\documentclass[12pt]{iopart}

\usepackage{graphicx}

\begin{document}

\title{$\mathcal{PT}$-symmetric optical superlattices}

\author{Stefano Longhi}

\address{Dipartimento di Fisica,
Politecnico di Milano and Istituto di Fotonica e Nanotecnologie del Consiglio Nazionale delle Ricerche, Piazza L. da Vinci 32, I-20133 Milano, Italy}
\ead{longhi@fisi.polimi.it}
\begin{abstract}
The spectral and localization properties of $\mathcal{PT}$-symmetric optical superlattices, either infinitely extended or truncated at one side, are theoretically investigated, and the criteria that ensure a real energy spectrum are derived. The analysis is applied to the case of superlattices describing a complex ($\mathcal{PT}$-symmetric) extension of the Harper Hamiltonian in the rational case.  
\end{abstract}

%Uncomment for PACS numbers title message
% Keywords required only for MST, PB, PMB, PM, JOA, JOB?
% non-Hermitian quantum mechanics; Bragg scattering; complex crystals;
%\vspace{2pc}
%\noindent{\it Keywords}: Article preparation, IOP journals
% Uncomment for Submitted to journal title message
%\submitto{\JPA}
% Comment out if separate title page not required

\maketitle

\section{Introduction}
Since the pioneering works by Bender and coworkers
 \cite{Bender1,Bender2,Bender3}, the properties of non-Hermitian Hamiltonians
that are symmetric with respect to combined parity ($\mathcal{P}$) and time-reversal ($\mathcal{T}$) operations have been extensively investigated by several authors (see, for instance, \cite{palle1,palle2,palle3,palle4,palle5,palle5bis,palle6,palle7} and references therein).  Although not Hermitian, $\mathcal{PT}$ symmetric Hamiltonians  may have purely real spectrum over a wide range of parameters in the so-called unbroken $\mathcal{PT}$ phase. $\mathcal{PT}$ symmetric Hamiltonians can be derived as effective models in many open quantum or classical systems, including matter wave \cite{mat1,mat2,mat3,mat4}, optical \cite{O2,O3,O4,O5,O6,O7,O8,Science,refe1,refe2,refe3,O9,O10,O11} and electronic \cite{electro1,electro2} systems. \par Among $\mathcal{PT}$ symmetric structures, Hamiltonians with a periodic and complex potential (so-called complex crystals) have received a great attention \cite{C1,C2,C3,C4,C4bis,C5,C6,C6tris,C6bis,C6quatris,C7,Berry,L1,L2,L3,L4,L5,L6,L7,L8} and recently found interesting realization in optics \cite{O9}. 
Complex crystals show rather unusual scattering and transport properties as compared to ordinary crystals, such as violation of the Friedel's law of Bragg
scattering \cite{C5,C6,L1}, double refraction and nonreciprocal diffraction \cite{O3}, unidirectional Bloch oscillations \cite{L4}, unidirectional invisibility \cite{L5,L6,L7,L8,L4bis}, 
and invisible defects \cite{L9,L10}. Complex crystals described by tight-binding Hamiltonians with complex site energies and/or hopping rates have been investigated in several recent works (see, for instance, \cite{palle5,palle5bis,palle6,palle7,O9,O11,L9,chain1,chain2,chain3,chain4,chain5,chain6,chain7} and references therein). Most of previous studies on non-Hermitian lattices have been limited to consider periodic or bi-periodic crystals,  inhomogenous lattices, or lattices in presence of localized defects or disorder. Superlattices, such as semiconductor and optical ones, provide an important class of synthetic crystals, where an additional periodicity is added on an underlying periodic structure. Tight binding superlattice models play an important role in several physical fields and can disclose a rich physics. They have been used, for instance, to realize  the Harper Hamiltonian \cite{Harper} and for the observation of exotic phenomena like the fractal energy spectrum of HofstadterÕs \cite{Hof} (see \cite{os1,os2,os3,os4,os5}).\par Motivated by such previous studies on $\mathcal{PT}$-symmetric Hamiltonians and superlattices, in this work we investigate theoretically the spectral properties of $\mathcal{PT}$-symmetric tight-binding optical superlattices. The general criterium for the existence of an entire real energy spectrum  (unbroken $\mathcal{PT}$ phase) for the infinitely extended lattice is derived, and the role of edge states in breaking the real energy spectrum is highlighted for the case of truncated lattices. As an example, we discuss the spectral and localization properties of $\mathcal{PT}$-symmetric superlattices that  realize a non-Hermitian extension of the Harper Hamiltonian in the rational case.

\section{Energy spectrum of $\mathcal{PT}$-symmetric optical superlattices}
Let us consider an optical superlattice described by a $\mathcal{PT}$-symmetric tight-binding Hamiltonian $\mathcal{H}$
\begin{equation}
\mathcal{H}_{n,m}= -\kappa_{n-1} \delta_{n,m+1}-\kappa_{n} \delta_{n,m-1}+V_n \delta_{n,m}
\end{equation}
where $\kappa_n$ is the hopping rate between lattice sites $n$ and $(n+1)$, which is assumed to be a real and non-vanishing number, and $V_n$ is the complex optical potential at site $n$. The Hamiltonian (1) can describe, for instance, light propagation along an array of optical waveguides with inhomogeneous waveguide spacings (that determine the hopping rates $\kappa_n$) and with engineered propagation constants and gain/loss terms in each waveguide (that determine the complex optical potential $V_n$); see for instance \cite{palle6}.  A superlattice with periodicity $q$ is obtained by assuming
\begin{eqnarray}
V_{n+q} & = & V_n \\
\kappa_{n+q} & = & \kappa_n
\end{eqnarray}
for some integer $q$. In an infinitely-extended lattice, the time reversal $\mathcal{T}$ and parity $\mathcal{P}$ operators are defined by $\mathcal{T}i \mathcal{T}=-i$ and 
$\mathcal{P}\hat{a}^{\dag}_n \mathcal{P}=\hat{a}^{\dag}_{-n}$, where $\hat{a}^{\dag}_n$ is the particle creation operator at lattice site $n$ \cite{chain2}. In practice, application of the $\mathcal{PT}$ operator to Eq.(1) changes $n \rightarrow -n$ and makes the complex conjugation of the complex potential. The infinitely-extended superlattice is $\mathcal{PT}$-symmetric provided that the additional constraints are satisfied
\begin{eqnarray}
\kappa_{-n} & = & \kappa_{n-1} \\
V_{-n} & = & V_n^*.
\end{eqnarray}
In this case, if $\psi_n$ is an eigenstate of $\mathcal{H}$ with energy $E$, i.e.
\begin{equation}
E \psi_n=-\kappa_{n-1} \psi_{n-1}-\kappa_n \psi_{n+1}+V_n \psi_n
\end{equation}
then $\psi_{-n}^*$ is an eigenstate of $\mathcal{H}$ with energy $E^*$. This implies that the energy spectrum of $\mathcal{H}$ is either entirely real or composed by pairs of complex conjugate numbers. In the former case the system is said to be in the unbroken $\mathcal{PT}$ phase.  Rather generally, the complex potential $V_n$ can be written as
\begin{equation}
V_n=V_n^{(R)}+i \lambda V_n^{(I)}
\end{equation}
where $V_n^{(R)}$ and $\lambda V_{n}^{(I)}$ are the real and imaginary parts of $V_n$, respectively, and $\lambda>0$ is a dimensionless parameter that measures the non-Hermitian  strength of the potential. For $\lambda=0$, $\mathcal{H}$ is Hermitian and the energy spectrum is entirely real-valued; as $\lambda$ is increased, a critical (threshold) value $\lambda_c \geq 0$ is found, above which pairs of complex energies will emerge (broken $\mathcal{PT}$ phase).

 In this section we would like to derive general criteria for an infinitely-extended optical superlattice to have an entirely real-valued energy spectrum. To this aim, let us notice that, according to the Bloch-Floquet theorem any  solution to Eq.(6) which does not diverge as $n \rightarrow \pm \infty$ can be taken to satisfy the constraint
\begin{equation}
\psi_{n+q} = \psi_{n} \exp(ikq)
\end{equation}
where $k$ is an arbitrary real parameter (the Bloch wave number), which varies in the range 
\begin{equation}
-\frac{\pi}{q} \leq k < \frac{\pi}{q}.
\end{equation}
Hence, if we write Eq.(6) for $n=1,2,...,q$ and use the relations $\psi_0=\psi_q \exp(-ikq)$ and $\psi_{q+1}=\psi_1 \exp(ikq)$ the following linear system of $q$ homogeneous equations in the $q$  amplitudes $\psi_1, \psi_2, ...., \psi_q$ is obtained
\begin{equation}
E \left(
\begin{array}{c}
\psi_1 \\
\psi_2 \\
\psi_3 \\
... \\
\psi_q
\end{array}
\right)= \mathcal{R}(k)
\left(
\begin{array}{c}
\psi_1 \\
\psi_2 \\
\psi_3 \\
... \\
\psi_q
\end{array}
\right)
\end{equation} 
where the $q \times q$ matrix $\mathcal{R}(k)$ is given by
\begin{equation}
\mathcal{R}(k) = \left( 
\begin{array}{cccccccc}
V_1 & -\kappa_1 & 0 & 0 & ... & 0 & 0 & -\kappa_{q} \exp(-ikq) \\
-\kappa_1 & V_2 & -\kappa_2 & 0 & ... & 0 & 0 & 0 \\
0 & -\kappa_2 & V_3 & -\kappa_3 & ...& 0 & 0 & 0 \\
... & ...& ... & ... & ... & ... & ...& ... \\
-\kappa_q \exp(ikq) & 0 & 0 & 0 & ... & 0 & -\kappa_{q-1} & V_q
\end{array}
\right) \;\;\;\;\;
\end{equation}
Equation (10) shows that the spectrum of $\mathcal{H}$ are the eigenvalues $E_1(k)$, $E_2(k)$, ..., $E_q(k)$ of the matrix $\mathcal{R}(k)$ for $k$ varying in the interval (9). Hence the superlattice sustains $q$ energy bands with dispersion relations $E_n(k)$ ($n=1,2,...,q$). It can be readily shown that $E_n(-k)=E_n(k)$, i.e. the real and imaginary parts of $E_n(k)$ are even functions of $k$. In the Hermitian case ($\lambda=0$), the energies are real-valued and the spectrum is formed by $q$ energy bands, separated by up to $(q-1)$ energy gaps.  As $\lambda$ is increased from $\lambda=0$, the gaps generally shrink until the symmetry breaking point is reached ($\lambda=\lambda_c$), at which  a gap disappears and an exceptional point emerges at either $k=0$ or $k= -\pi/ q$ \cite{L1}.  Therefore, the following general criterium can be stated:\par
{\it Theorem I}. The  Hamiltonian (1) for an infinitely-extended superlattice with period $q$ has an entirely real-valued energy spectrum (i.e. it is in the unbroken $\mathcal{PT}$ phase) if and only if the $ 2\times q$ eigenvalues of the $q \times q$ matrices $\mathcal{R}_1=\mathcal{R}(0)$ and $\mathcal{R}_2=\mathcal{R}(-\pi / q)$, defined by Eq.(11), are real. 

\section{Semi-infinite superlattice and edge states}
Lattice truncation at one side generally introduces edge (surface) states, which can be associated to complex energies even though the infinitely-extended lattice is in the unbroken $\mathcal{PT}$ phase. To study the emergence of edge states, let us assume that the lattice is truncated at the left side (semi-infinite lattice), i.e. the lattice sites are $n=1,2,3,....$.  In this case, the definition of the parity operator $\mathcal{P}$ becomes meaningless, and hence for the semi-infinite lattice $\mathcal{PT}$ invariance of the Hamiltonian can not be applied and eigen-energies do not necessarily emerge in complex conjugate pairs. According to Eq.(6), for an eigenstate $(\psi_1, \psi_2, \psi_3, .....)$ of $\mathcal{H}$ one can write
\begin{equation}
\left(
\begin{array}{c}
\psi_{n+1} \\
\psi_n
\end{array}
\right)
=\mathcal{M}_n (E) \left(
\begin{array}{c}
\psi_{n} \\
\psi_{n-1}
\end{array}
\right)
\end{equation}
where
\begin{equation}
\mathcal{M}_n(E)=
\left(
\begin{array}{cc}
(V_n-E)/ \kappa_n  & - \kappa_{n-1} / \kappa_n \\
1 & 0
\end{array}
\right).
\end{equation}
Taking into account that $\mathcal{M}_{n+q}(E)=\mathcal{M}_{n}(E)$, for any arbitrary integer number $M=0,1,2,3,4,...$ one then obtains
\begin{equation}
\left(
\begin{array}{c}
\psi_{Mq+1} \\
\psi_{Mq}
\end{array}
\right)
=\mathcal{S}^M (E) \left(
\begin{array}{c}
\psi_{1} \\
\psi_{0}
\end{array}
\right)
\end{equation}
where we have set
\begin{equation}
\mathcal{S}(E)=\mathcal{M}_q(E) \times \mathcal{M}_{q-1}(E) \times ... \times \mathcal{M}_1(E).
\end{equation}
From Eqs.(13) and (15) it can be readily shown that $\mathcal{S}_{11}(E)$, $\mathcal{S}_{12}(E)$, $\mathcal{S}_{21}(E)$ and $\mathcal{S}_{22}(E)$ are polynomials of $E$ of order $q$, $q-1$, $q-1$ and $q-2$, respectively. Moreover $det \mathcal{S}=\mathcal{S}_{11} \mathcal{S}_{22}-\mathcal{S}_{12} \mathcal{S}_{21}=1$. Since the $ 2 \times 2$ matrix $\mathcal{S}(E)$ is unimodular, the $M$-power of $\mathcal{S}(E)$ can be calculated in a closed form and reads (see, for instance, \cite{note})
\begin{equation}
\mathcal{S}^M(E)= \frac{1}{\sin \theta} 
\left(
\begin{array}{cc}
\mathcal{S}_{11} \sin (M \theta)- \sin [(M-1) \theta] & \mathcal{S}_{12} \sin (M \theta) \\
\mathcal{S}_{21} \sin (M \theta) & \mathcal{S}_{22} \sin \theta - \sin [(M-1) \theta]
\end{array}
\right)
\end{equation}
where the angle $\theta$ is defined by the relation
\begin{equation}
\cos \theta= \frac{\mathcal{S}_{11}+\mathcal{S}_{22}}{2}.
\end{equation}
For the semi-infinite lattice truncated at the lattice site $n=1$, the boundary condition {$\psi_0=0$ applies. Once the (non-vanishing) amplitude $\psi_1$ has been assigned, Eqs.(14) and (16) allow one to calculate the modal amplitudes $\psi_{Mq}, \psi_{Mq+1}$ ($M=1,2,3,...$) all along the lattice. Without loss of generality, for our purposes we may assume $\psi_1=1$; a different choice of $\psi_1$ just results in a multiplication of the lattice eigenmode by a constant.  With $\psi_0=0$ and $\psi_1=1$}, from Eqs.(14) and (16)   
one then obtains
\begin{eqnarray}
\psi_{Mq+1} & = &  \frac{\mathcal{S}_{11} \sin (M \theta)- \sin [(M-1) \theta] }{\sin \theta} \\
\psi_{Mq} & = & \frac{\mathcal{S}_{21} \sin (M \theta)}{\sin \theta} 
\end{eqnarray}
for $M=1,2,3,4,....$. The energy spectrum of the Hamiltonian $\mathcal{H}$ for the semi-infinite lattice is provided by the values of $E$ such that the sequences $\psi_{Mq}$, $\psi_{Mq+1}$ do not diverge as $M \rightarrow \infty$. As shown in the Appendix, this can occur in two cases solely:\\
(i) The angle $\theta$ is a real number, i.e. $\mathcal{S}_{11}+\mathcal{S}_{22}$  is real \cite{nota2} and 
\begin{equation}
| \mathcal{S}_{11}+\mathcal{S}_{22}|<2.
\end{equation}
This case is satisfied for any value $E$ belonging to the energy spectrum of the infinitely-extended lattice and the corresponding eigenstates correspond to scattered (non-normalizable) states of the semi-infinite superlattice (see the Appendix).\\
(ii) $E$ satisfies the condition
\begin{equation}
\mathcal{S}_{21}(E)=0 \;  {\rm with}  \; \; |\mathcal{S}_{11} (E) | \leq 1.
\end{equation}
In this case the eigenstate is an edge (normalizable) state if  $|\mathcal{S}_{11} (E) | < 1$, whereas it is a scattered (extended) state if $ |\mathcal{S}_{11} (E) |=1$.  Note that, for $|\mathcal{S}_{11} (E) | < 1$, the edge state shows an exponential localization with a localization length $L= 
-q / {\rm ln} | \mathcal{S}_{11}|^2 $ (see the Appendix). Since $S_{21}(E)$ is a polynomial in $E$ of order $(q-1)$, the equation $\mathcal{S}_{21}(E_0)=0$ is satisfied by $(q-1)$ complex numbers $E_1$, $E_2$,..., $E_{q-1}$.  The roots $E_l$ of the equation $\mathcal{S}_{21}(E_l)=0$ with $|\mathcal{S}_{11}(E_l)|=1$, corresponding to extended states, belong to the continuous spectrum; as shown in the Appendix, if the infinitely-extended lattice is in the unbroken $\mathcal{PT}$ phase, $E_l$ are real.\\
The two above properties show that the energy spectrum of the semi-infinite lattice differs from that of the infinite lattice, considered in the previous section, for up to $(q-1)$ additional energies, corresponding to edge (normalizable) states. The following general criterium for the reality of the energy spectrum of a semi-infinite superlattice can be thus stated:\par
{\it Theorem II.} The energy spectrum of the Hamiltonian (1) for a semi-inifinite superlattice comprising the lattice sites $n=1,2,3,4,....$ is entirely real if and only if the criterium stated in Theorem I is satisfied (i.e. the infinitely-extended lattice is in the unbroken $\mathcal{PT}$ phase) and the energies $E_l$ of edge states, obtained from the algebraic equation $\mathcal{S}_{21}(E_l)=0$ with $|\mathcal{S}_{11}(E_l)| < 1$, are real numbers. Here $\mathcal{S}_{21}(E)$ and $\mathcal{S}_{11}(E)$ are the coefficients of the $2 \times 2$ matrix $\mathcal{S}$ defined by Eqs.(13) and (15).\\
\\
From a computational viewpoint, the roots $E_l$  ($l=1,2,...,q-1$) of the equation $\mathcal{S}_{21}(E)=0$ can be determined in an easier way as follows. Since at $E=E_l$  $\psi_{n}$ vanishes at $n=0$ ($\psi_0=0$) and at $n=q$ ($\psi_q=0$), from Eq.(6) one has
\begin{equation}
E_l \left(
\begin{array}{c}
\psi_1 \\
\psi_2 \\
... \\
\psi_{q-1}
\end{array} 
\right)= \mathcal{Q}  \left(
\begin{array}{c}
\psi_1 \\
\psi_2 \\
... \\
\psi_{q-1}
\end{array} 
\right)
\end{equation}
where we have set
\begin{equation}
\mathcal{Q}= \left( 
\begin{array}{ccccccc}
V_1 & -\kappa_1 & 0 &  ...& 0  & 0 & 0  \\
-\kappa_1 & V_2 & -\kappa_2 &  ... & 0  & 0 & 0  \\
... & ...& ... & ... & ... & ... & ... \\
0 & 0 & 0 &  ... & 0 & -\kappa_{q-2} & V_{q-1}
\end{array}
\right) \;\;\;\;\;
\end{equation}
Hence $E_l$ can be computed as the eigenvalues of the $(q-1) \times (q-1)$ matrix $\mathcal{Q}$, defined by Eq.(23). Edge (surface) states corresponds to the eigenvalues $E_l$ of $\mathcal{Q}$ satisfying the constraint $| \mathcal{S}_{11}(E_l)|<1$. Hence the semi-infinite superlattice can sustain up to $(q-1)$ edge states.
\begin{figure}[htb]
\centerline{\includegraphics[width=10cm]{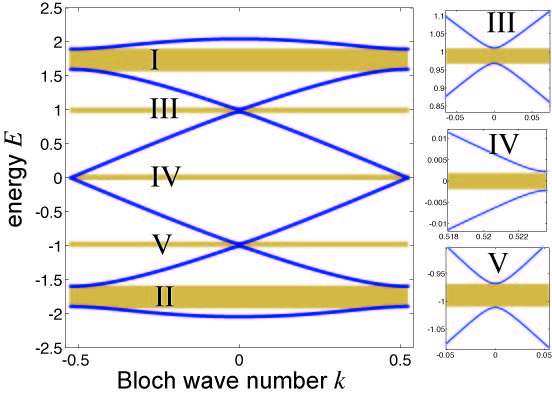}} \caption{
Numerically-computed energy spectrum of the Harper superlattice in the Hermitian case. Parameter values are $\delta=0.3$, $\lambda=0$, $p=1$ and $q=6$. The energy spectrum comprises 6 bands, separated by two wide energy gaps (indicated by I and II in the figure) and three narrow gaps (indicated by III, IV and V). The shaded regions are the energy gaps. The right panels show an enlargement of the narrow energy gaps III, IV and V.}
\end{figure}
\begin{figure}[htb]
\centerline{\includegraphics[width=10cm]{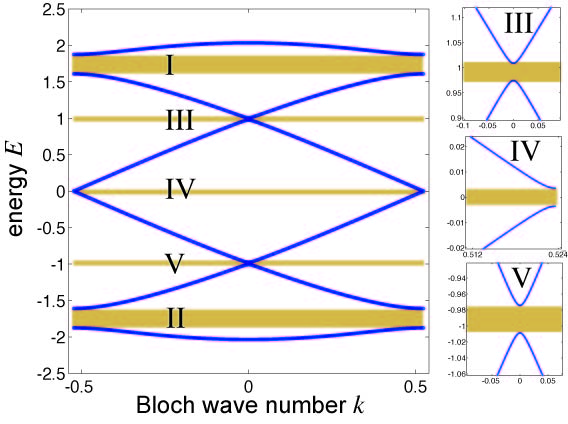}} \caption{
Same as Fig.1, but for $\lambda=0.134$.}
\end{figure}
\begin{figure}[htb]
\centerline{\includegraphics[width=10cm]{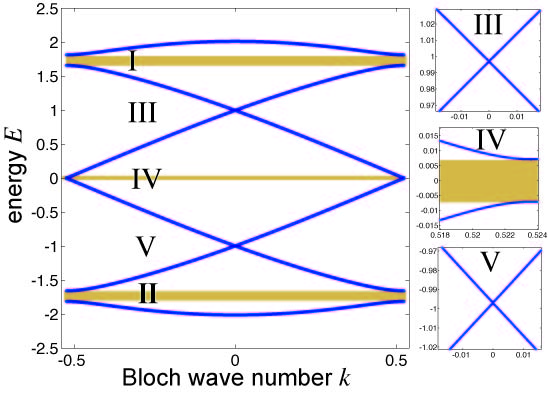}} \caption{
Same as Fig.1, but for $\lambda$ at the symmetry breaking point ($\lambda=\lambda_c=0.2552$). Note that at the symmetry breaking point the narrow gaps III and V vanish and the bands touch at $k=0$. At larger values of $\lambda$ complex-conjugate energies emerge near $k=0$.}
\end{figure}

\section{ $\mathcal{PT}$-symmetric Harper model}
As an application of the general analysis developed in the previous section, we consider a non-Hermitian ($\mathcal{PT}$ symmetric) extension of the famous Harper model in the rational case. The $\mathcal{PT}$ symmetric Harper model is obtained by assuming uniform hopping rates $\kappa_n=\kappa$ and a sinusoidal potential, namely
\begin{equation}
V_n=\delta \cos[2 \pi \alpha (n-n_0)]+i \lambda  \sin [2 \pi \alpha (n-n_0)]
\end{equation} 
where $\delta$ and $\lambda$ are real numbers and $n_0$ is a reference index. Without loss of generality, in the following we will assume $\kappa_n=\kappa=1$. The $\mathcal{PT}$-symmetric extension of the Harper equation reads
\begin{equation}
E \psi_n=\psi_{n+1}+\psi_{n-1}+\left\{ \delta \cos[2 \pi \alpha (n-n_0)] +i \lambda \sin[2 \pi \alpha(n-n_0)] \right\} \psi_n
\end{equation}
The usual (Hermitian) Harper model \cite{Harper,Hof}, which arises in the study of the quantum Hall effect and also known as the Aubry-Andre model \cite{Aubry}, is obtained by taking $\lambda=0$. In mathematical physics, $\mathcal{H}$ is also known as the quasi-Mathieu operator. The spectrum of the quasi-Mathieu operator is known to depend on whether $\alpha$ is a rational or irrational number. Here we consider the rational case, $\alpha=p/q$ with $p$ and $q$ irreducible integer numbers, so that the potential is periodic and the Harper model describes a superlattice.  
\subsection{Infinitely-extended Harper superlattice}
For an infinitely-extended superlattice, according to the results of Sec.2 the energy spectrum is composed by $q$ bands whose dispersion curves are obtained from the $q$ eigenvalues of the  matrix $\mathcal{R}(k)$ defined by Eq.(11), where $-\pi/q \leq k < \pi/q$ is the Bloch wave number. As an example, in Figs.1,2 and 3 we show the numerically-computed energy spectrum for $\delta=0.3$, $p=1$, $q=6$ and for a few increasing values of the non-Hermitian parameter $\lambda$ below the $\mathcal{PT}$ symmetry breaking point \cite{note3}.  In the Hermitian case ($\lambda=0$, see Fig.1) the superlattice sustains 6 bands, separated by two wide gaps I and II and three small gaps III, IV and V, see Fig.1. As $\lambda$ is increased, the energy spectrum remains entirely real valued, however the small gaps III and V get narrower, see Fig.2. At the symmetry-breaking point ($\lambda=\lambda_c=0.2252$) the gaps III and V vanish and the corresponding bands touch at $k=0$, see Fig.3.\par 
The symmetry breaking threshold $\lambda_c$ turns out to depend sensitively on the ratio $\alpha=p/q$, and is found to be smaller than $\delta$. As an example, in Fig.4(a) we show the numerically-computed critical value $\lambda_c$ for $\delta=0.3$, $p=1$ and for increasing values of $q$ \cite{note4}. Note that the critical value $\lambda_c$ is always smaller than $\delta$, and vanishes for $q=4,8,12,16,20,...$. As $q$ is increased, the local maximum of $\lambda_c$ gets smaller. However, for a fixed value of $\lambda$ above the critical value $\lambda_c$, the maximum growth rate of the most unstable eigenenergy, defined as ${\sigma=\rm max} \{ {\rm Im} (E) \}$, is found to decrease as $q$ increases, with an almost exponential decay; see Fig.4(b). Form a physical viewpoint this means that, while $\mathcal{PT}$ symmetry gets fragile as the superlattice period $q$ becomes large, the growth rate of the unstable states at a given value of $\lambda$ above the symmetry breaking threshold $\lambda_c$ exponentially diminishes with $q$.\\ 
In Fig.5 we show the behavior of $\lambda_c$ and $\sigma$ for a fixed value of $q=19$ (prime number), $\delta=0.3$ and for increasing values of $p$, from $p=1$ to $p=q-1=18$. Interestingly, in this case above the $\mathcal{PT}$ symmetry breaking threshold the maximum growth rate $\sigma$ at a given value of $\lambda> \lambda_c$ ($\lambda=\delta$ in the figure) is almost independent of $p$; see Fig.5(b).

\begin{figure}[htb]
\centerline{\includegraphics[width=13cm]{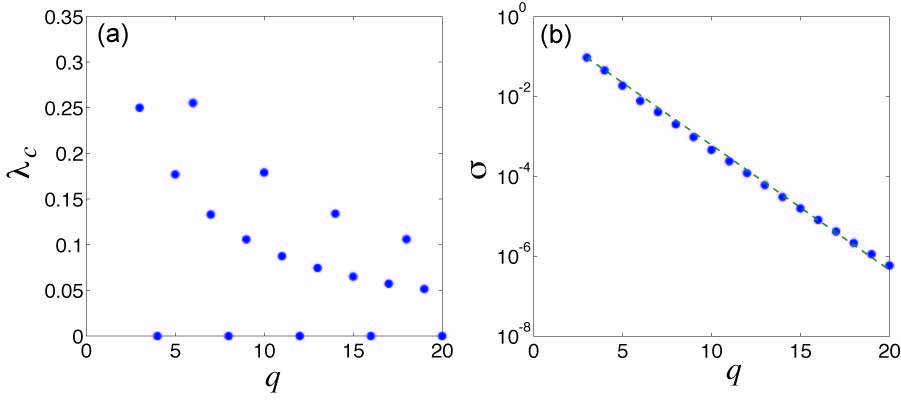}} \caption{
(a) Numerically-computed critical value $\lambda_c$ of $\mathcal{PT}$ symmetry breaking versus $q$ for the Harper superlattice for parameter values $\delta=0.3$ and $p=1$. (b) Behavior of the maximum growth rate $\sigma$ versus $q$ above the $\mathcal{PT}$ symmetry breaking for $\delta=0.3$ and $\lambda=\delta$.The fitting dashed curve is the exponential function $\sigma=0.8221 \times \exp(-0.72q)$.}
\end{figure}
\begin{figure}[htb]
\centerline{\includegraphics[width=13cm]{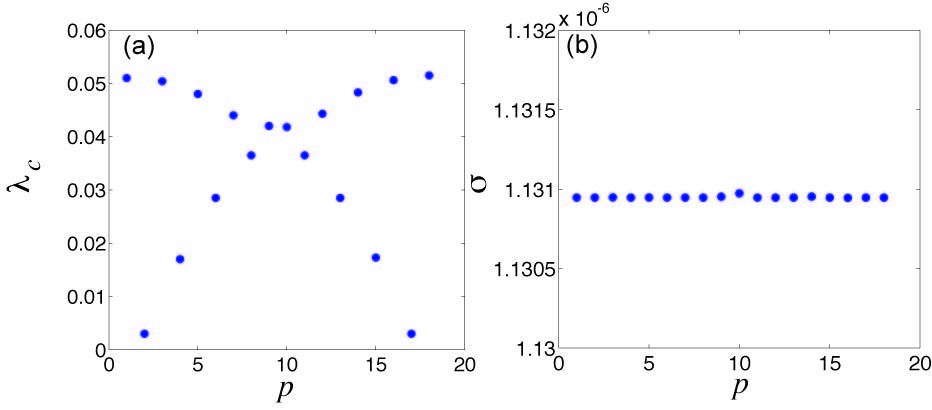}} \caption{
Same as Fig.4, but for $q=19$ and $\delta=0.3$. In (b) $\lambda=\delta=0.3$.}
\end{figure}

\subsection{Semi-infinite Harper superlattice}
In the semi-infinite Harper superlattice, according to the Theorem II of Sec.3 complex energies can arise owing to the emergence of edge states. Let us consider an infinitely-extended Harper superlattice which is in the unbroken $\mathcal{PT}$ phase, i.e. for $\lambda < \lambda_c$. Lattice truncation at site $n=1$ can introduce up to $(q-1)$   
allowed energies $E$, which are obtained as the roots of the algebraic equation (of order $q-1$) $\mathcal{S}_{21}(E)=0$ satisfying the constraint $|\mathcal{S}_{11}(E)| \leq 1$. Edge (surface) states correspond to $|\mathcal{S}_{11}(E)| < 1$, whereas $|\mathcal{S}_{11}(E)| = 1$ yields scattered states.
 Extended numerical simulations show quite generally that the emergence of edge states is associated to the appearance of complex energies below the critical value $\lambda_c$. Conversely, semi-infinite lattices that do not sustain edge states maintain a real energy spectrum like the infinitely-extended lattice. As an example, let us consider the Harper superlattice with $p=1$, $q=6$, $\delta=0.3$ and $\lambda=0.134<\lambda_c$. The infinitely-extended superlattice is in the unbroken $\mathcal{PT}$ phase and its energy spectrum was shown in Fig.2. Let us now consider the truncated lattice, and let us compute the roots of the  equation $\mathcal{S}_{21}(E)=0$, satisfying the constraint $|\mathcal{S}_{11}(E)| \leq 1$, for a few values of the reference index $n_0$ [see Eq.(24)]. Changing the value of $n_0$ physically means that we cut the lattice at different positions of  the superlattice period. The results are summarized in Table 1. 
\Table{\label{tabl1} Edge states and corresponding energies in the semi-infinite Harper superlattice for $p=1$, $q=6$, $\delta=0.3$, $\lambda=0.134$, and for a few values of the reference index $n_0$. For each value of $n_0$, the energies $E_l$ shown in the table are the roots of the algebraic equation $\mathcal{S}_{21}(E_l)=0$ satisfying the constraint $|\mathcal{S}_{11}(E_l)| \leq 1$. Edge states correspond to  $|\mathcal{S}_{11}(E_l)| < 1$, extended states to  $|\mathcal{S}_{11}(E_l)| =1$.}
\br
%&&&\centre{2}{uff}\\
%\ns
%&Thickness&&\crule{2}\\
$n_0$ & spectrum & energy $E_l$ & $|\mathcal{S}_{11}(E_l)|$ &\\
\mr
0 & real & -1.8850 & 1 & extended state \\
   &                    &  -1.0147 & 1 & extended state \\
     &                 & -0.0036  & 1 & extended state \\
     &                 & 1.0233   & 1 & extended state \\
     &                 & 1.5799  & 1 & extended state \\
\mr
1 & complex & 1.7058 + 0.0712i & 0.4322 & edge state\\     
\mr
2 & complex & 1.8413 + 0.0410i & 0.4989 & edge state\\
   &               &   0.9872 + 0.0166i & 0.9199 & edge state \\
    &              &  -0.0034 - 0.0001i & 0.9968 & edge state \\
      &            &  -0.9693 - 0.0126i & 0.9449 & edge state \\
\mr
3 & real & -1.5799 & 1 & extended state \\
  &    &  -1.0233  & 1 & extended state \\
   & & 0.0036 & 1 & extended state \\
   & & 1.0147 & 1 & extended state \\
    & & 1.8850 & 1 & extended state \\
    \mr
    4 & complex & -1.7058 - 0.0712i & 0.4322 & edge state \\
    \mr 
    5 & complex & -1.8413 - 0.0410i & 0.4989 & edge state \\
     & & -0.9872 - 0.0166i & 0.9199 & edge state \\
     & & 0.0034 + 0.0001i & 0.9968 & edge state \\
     & & 0.9693 + 0.0126i & 0.9449 & edge state \\
    \br
\end{tabular}
\end{indented}
\end{table}
Note that, for $n_0=0$ and $n_0=3$, there are no edge states and the energies $E_l$ of extended states are real and are located at band edges. Hence the truncated superlattice maintains  an entire real energy spectrum. Conversely, for $n_0=1,2,4$ and 5 edge states with complex energies are found  (one edge state for $n_0=1,4$ and four edge states for $n_0=2,5$, see table 1). The physical reason why edge states are generally associated to complex energies is that, contrary to extended (scattered) states,  the exponentially-localized surface modes do not experience in a balanced manner the influence of positive and negative imaginary parts (i.e. optical gain and loss) of the complex potential, resulting in an effective dominance of either gain or loss. Such a behavior turns out to be rather insensitive to the value of $\lambda$, i.e. edge states are rather generally associated to complex energies. As an example, in Fig.6. we the behavior of the real and imaginary parts of the energies $E_l$ of edge states for lattice truncation at $n_0=1$ and $n_0=2$. Note that for $n_0=2$ an edge state with an almost vanishing imaginary part of the energy is found, however the other surface states are associated to complex energies.
\begin{figure}[htb]
\centerline{\includegraphics[width=12cm]{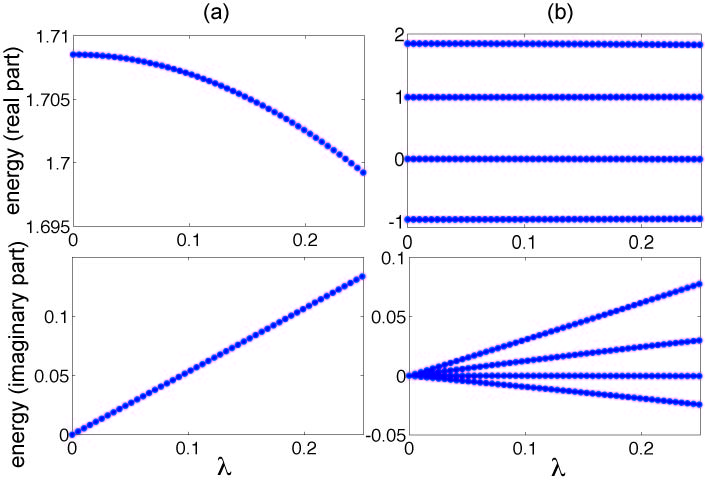}} \caption{
Behavior of the complex energies $E_l$ (real and imaginary parts) of edge states as a function of $\lambda$ for $\delta=0.3$, $p=1$, $q=6$, and for (a) $n_0=1$ and (b) $n_0=2$. In (a) there is only one edge state, whereas in (b) there are four edge states, one of which with an energy $E_l \simeq 0$.}
\end{figure}

\par
To highlight the role of edge states, in Fig.7 we show the evolution of the light intensity distributions $|\psi_n(t)|^2$ along the semi-infinite Harper superlattice for single-site excitation of the left boundary site $n=1$, i.e. for the initial condition $\psi_n(0)=\delta_{n,1}$. The maps shown in the figure have been obtained by numerical solution of the Schr\"{o}dinger equation 
\begin{equation}
i \frac{ \partial \psi_n}{\partial t}= \sum_m \mathcal{H}_{n,m} \psi_m= \psi_{n+1}+\psi_{n-1}+V_n \psi_n
\end{equation} 
  using an accurate fourth-order variable step Runge-Kutta method. Note that for $n_0=0$ [Fig.7(a)] there are not edge states and light does not remain localized near the lattice edge, undergoing discrete diffraction in the array. A similar behavior is found for $n_0=3$ (not shown in the figure). For $n_0=1$ and $n_0=2$, according to Table 1 there are edge states with complex energies and imaginary positive part. Hence light remains partially localized at the lattice edge and it is exponentially amplified [see Figs.7(b) and (c)]. For $n_0=4$, the semi-lattice sustains one edge state with complex energy and {\it negative} imaginary part (see table 1). This means that light does not remain localized at the lattice edge because the surface mode is exponentially damped (rather than amplified). This case is shown in Fig.7(d).
\begin{figure}[htb]
\centerline{\includegraphics[width=10cm]{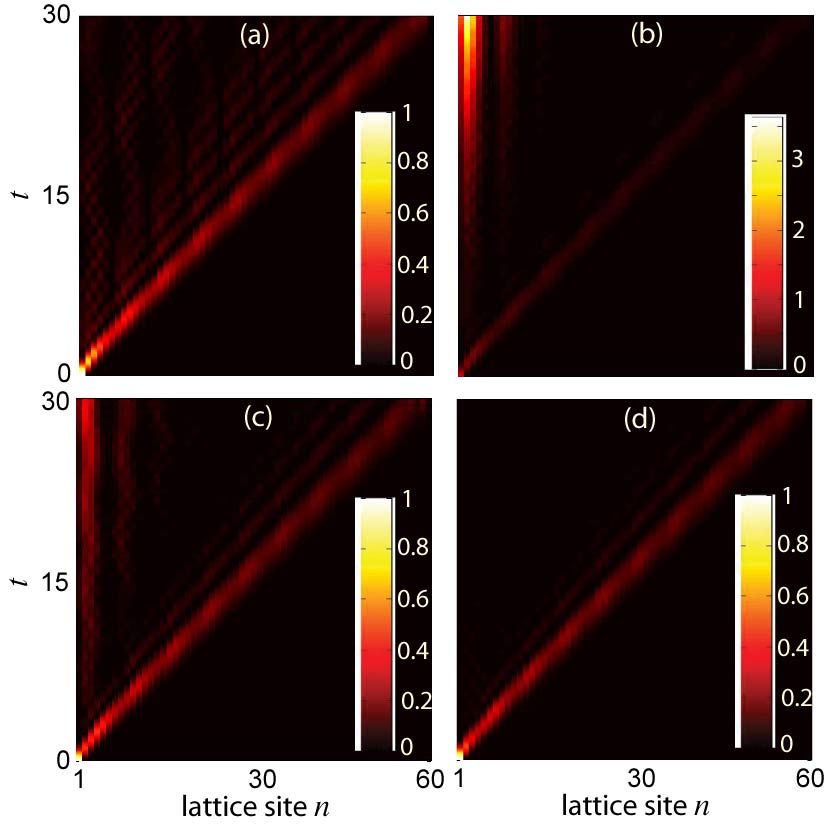}}
 \caption{Numerically-computed evolution of $|\psi_n(t)|^2$ in a semi-infinite Harper superlattice for $p=1$, $q=6$, $\delta=0.3$, $\lambda=0.134$ and for (a) $n_0=0$, (b) $n_0=1$, (c) $n_0=2$, and (d) $n_0=4$. The initial condition is $\psi_n(0)=\delta_{n,1}$ (excitation of the left-edge waveguide).}
\end{figure}

\section{Conclusions}
In this work we have investigate theoretically the spectral and localization properties of $\mathcal{PT}$-symmetric tight-binding optical superlattices, and we have derived the general criterium for the existence of an entire real energy spectrum  (unbroken $\mathcal{PT}$ phase). The role of edge states in the appearance of complex energies has been discussed for a semi-infinite lattice. The general analysis has been applied to study the spectrum and localization of a $\mathcal{PT}$-symmetric extension of the Harper model in the rational case. Interestingly, we found that the $\mathcal{PT}$ symmetry gets fragile as the superlattice period $q$ becomes large, however the growth rate of the unstable states at a given value of $\lambda$ above the symmetry breaking threshold $\lambda_c$ exponentially diminishes with $q$.  Our results could stimulate further theoretical and experimental investigations of the spectral, topological and localization properties of $\mathcal{PT}$-symmetric superlattices and quasi-crystals. For example, it would be interesting to investigate the topoloogical invariant properties of a gapped superlattice (e.g. the Chern numbers) in the complex (non-Hermitian) case, or the spectral properties of the $\mathcal{PT}$-symmetric Harper model in the quasi-crystal (i.e. for an irrational value of $\alpha$) limit \cite{kazzz}.

\appendix
\section{Energy spectrum of the semi-infinite optical superlattice}

The spectrum of the semi-infinite superlattice corresponds to the energy values $E$ such that the solution $\psi_n$ to Eq.(6) is a normalizable state ($\sum_{n=1}^{\infty} | \psi_n|^2 < \infty$, point spectrum of $\mathcal{H}$) or it is a non-normalizable but limited function as $n \rightarrow \infty$ (continuous spectrum of $\mathcal{H}$). Since the superlattice is periodic with period $q$, the above conditions can be applied to $\psi_{Mq}$ ($M=1,2,3,....$) rather than to $\psi_n$.\par Let us first assume $\mathcal{S}_{21}(E) \neq 0$. Form Eq.(19) it then readily follows that, if the angle $\theta$ [defined by Eq.(17)] is a complex number (with non-vanishing imaginary part), the solution $\psi_{Mq}$ is unbounded as $M \rightarrow \infty$, and hence $E$ does not belong to the spectrum of $\mathcal{H}$. Conversely, for values of $E$ such as $\theta$ is real, from Eq.(19) is follows that $\psi_{Mq}$ is an oscillating (limited) function of $M$ as $M \rightarrow \infty$, i.e. $E$ belongs to the continuous spectrum of $\mathcal{H}$. It can be readily shown that the range of values of $E$ for which the angle $\theta$ is real corresponds to the continuous spectrum of the infinitely-extended superlattice. To prove this statement, let us consider the infinitely-extended superlattice ad let us indicate by $ \psi_n(k)$ the Bloch-Floquet eigenstate with wave number $k$ and energy $E$ belonging to one band of the superlattice. Since $\psi_{q}=\psi_0 \exp(ikq)$ and $\psi_{q+1}=\psi_1 \exp(ikq)$, from Eq.(14) with $M=1$ one has
\begin{equation}
\exp(ikq) \left(
\begin{array}{c}
\psi_{1} \\
\psi_{0}
\end{array}
\right)
=\mathcal{S} (E) \left(
\begin{array}{c}
\psi_{1} \\
\psi_{0}
\end{array}
\right)
\end{equation}
i.e. $\exp(ikq)$ is an eigenvalue of $\mathcal{S}(E)$. On the other hand, the eigenvalues of $\mathcal{S}(E)$ are the roots of the second-order algebraic equation 
$\lambda^2-(\mathcal{S}_{11}+\mathcal{S}_{22}) \lambda+1=0$, which read $\lambda_{\pm}=\exp( \pm i \theta)$. Hence it follows that $\cos \theta=\cos(kq)$, i.e. the angle $\theta$ is real. This proves that, if $E$ belongs to the continuous spectrum of the infinitely-extended superlattice, then it also belongs to the continuous spectrum of the semi-infinite lattice. Conversely, let us assume that $E$ does not belong to the continuous spectrum of the infinitely-extended superlattice. In this case, according to the Bloch-Floquet theorem the solutions to Eq.(6) can be chosen to satisfy the condition $\psi_{n+q}= \pm \psi_n \exp(\mu q)$, where $\mu$ is a real and non-vanishing number. In this case one would obtain 
$\pm \exp(\mu q)=\exp( \pm i \theta)$, i.e. $\theta$ would have a non-vanishing imaginary part and thus $E$ does not belong to the continuous spectrum of the semi-infinite lattice.\par
Let us now consider the case $\mathcal{S}_{21}(E)=0$. Since $\mathcal{S}_{21}(E)$ is a polynomial in $E$ of order $(q-1)$, the equation  $\mathcal{S}_{21}(E)=0$ in the complex $E$ plane is satisfied for $(q-1)$ values $E=E_1$, $E=E_2$, ... , $E=E_{q-1}$.  In this case, from Eqs.(18) and (19) one obtains $\psi_{Mq}=0$ and $\psi_{Mq+1}=\mathcal{S}_{11}^M(E)$ for $M=0,1,2,3,...$. Hence, for $l=1,2,...,q-1$:\\
(i)  If $|\mathcal{S}_{11}(E_l)|<1$, $\psi_n$ is an exponentially-localized edge state and $E_l$ belongs to the point spectrum of the semi-infinite lattice.\\
(ii) If $|\mathcal{S}_{11}(E_l)|=1$, $\psi_n$ is an extended but limited function of $n$ and $E_l$ belongs to the continuous spectrum.\\ 
(iii) If  $|\mathcal{S}_{11}(E_l)|>1$, $\psi_n$ is not a limited function as $n \rightarrow \infty$ and  $E_l$ does not belong to the spectrum of the semi-infinite superlattice.\\
Note that, in case (i), i.e. if the energy $E_l$ corresponds to an edge (surface) state, since $\psi_{Mq+1}= \mathcal{S}_{11}^M$ ($M=0,1,2,3,...$) it follows that the state $\psi_n$ shows an exponential localization with a localization length 
\begin{equation}
L=-\frac{q}{ {\rm ln} | \mathcal{S}_{11}|^2}.
\end{equation}


\begin{thebibliography}{10}
\bibitem{nota2}
For $E$ real, the trace $\mathcal{S}_{11}+\mathcal{S}_{22}$ of $\mathcal{S}(E)$ is real.
\bibitem{note3}
For the infinitely-extended superlattice, the spectrum does not depend on the reference index $n_0$. However, for the semi-infinite lattice edge states will sensitively depend on the value of $n_0$; see Table 1.
\bibitem{note4}
The case $p=1$ and $q=2$, corresponding to a bipartite lattice, was investigated in previous works (see, for instance, \cite{L3}). In this case $\lambda_c=0$.
\end{thebibliography}
\end{document}